\documentclass[aps,prab,reprint,superscriptaddress,showpacs]{revtex4-1}

\usepackage{graphicx}
\usepackage{amsmath, amssymb, amsfonts}
\usepackage[usenames]{color}
\usepackage{soul}

\begin{document}
	
	\title{Radiation of a charge moving along a corrugated surface with a relatively small period}
	
	\author{Evgeniy S. Simakov}
	
	\author{Andrey V. Tyukhtin}
	
	\author{Sergey N. Galyamin}
	
	\affiliation{Saint Petersburg State University, 7/9 Universitetskaya nab., St. Petersburg, 199034 Russia}
	
	\date{\today}
	
\begin{abstract}
We consider electromagnetic radiation of a charged particle bunch moving uniformly along a corrugated planar metallic surface. 
It is assumed that the wavelengths under consideration are much larger than the period and the depth of corrugation. 
Using the method of the equivalent boundary conditions we obtain the Fourier-transform of the Hertz vector. 
It is demonstrated that the ultra-relativistic bunch excites the surface waves, whereas the volume radiation is absent. 
Fourier-transforms of the surface wave components and spectral density of energy losses are obtained and analyzed.
\end{abstract}


\maketitle

\section{Introduction}
Electromagnetic interaction of charged particle beams with corrugated structures is of essential interest nowadays for a number of reasons. For example, prospective applications of these structures are connected with microwave and Terahertz (THz) emission. 
One should mention here a theoretical treatment (based on surface impedance formalism) of bunch wakefields generated in flat and circular metallic waveguides with shallow (fine) corrugation~%
\cite{StupBane12, BanStup16, BaneStupakovZagorodnov16, TVAA2018}.
Possibilities for microwave and THz radiation in such structures were also studied, including simulations~%
\cite{BanStup12THz}
and experiments~%
\cite{Yalandin97, GinzburgNS16, BanStupAntipov17, He17, MesyatsGinzburg17}. 
Interaction between continuous electron beam and corrugated surface was analyzed in papers~%
\cite{Mineo10, GinzburgNS13, GinzburgMalkin16}
in view of realization of relativistic surface-wave amplifiers.

Concerning the radiation generated by charged particle bunches moving along periodic structures, one should mention extended researches of various cases where wavelengths are comparable or less than the structure period. 
This type of radiation is usually referred to as Smith-Purcell radiation, and there are plenty of the literature devoted to this subject (see, for example, books~%
\cite{PotRyazTishchenkob11,Potb11}%
).

In this paper, we study theoretically the situation, which is principally different from Smith-Purcell case: the wavelengths under consideration are supposed to be much larger than periods of the structure, i.e. we consider so-called ``long-wave'' (or ``low-frequency'') approximation. 
In this case, the periodic structure can be described by so-called equivalent boundary conditions (EBC), which should be fulfilled on the flat surface (they are also known as Vainstein-Sivov conditions)~%
\cite{NefSivb77}. 

Contrary to the aforementioned papers dealing with waveguide structures with shallow corrugation~%
\cite{StupBane12, BanStup16, BaneStupakovZagorodnov16, TVAA2018}%
, here we analyze the case of a point charge (or \textbf{a} one-dimensional finite gaussian bunch) moving along infinite flat corrugated surface.
This problem has not been solved previously, despite numerous papers dedicated to radiation from charged particle bunches moving in the presence of periodic corrugated structures.

It should be underlined that the paper~%
\cite{TVAA2018} 
contains the comparison between theoretical results (obtained using the EBC approach) and results of the CST Particle Studio simulations. 
The authors have demonstrated that the ultra-relativistic Gaussian bunch excites monochromatic wakefield, with the simulated wavelength being equal approximately to $10$ mm (for the period of corrugation $1$ mm). 
The difference between theoretical and simulated wavelengths varies from 
$0.2\%$ 
to 
$4\%$ 
depending on the width of ``hills'' of the corrugated structure. 
The coincidence can be considered as good, and this circumstance justifies the use of the EBC method even in situations where the structure period is only several times shorter than the considered wavelengths.

It is also notable that the problem under consideration is close to similar problems where the periodic structure is not solid. 
In particular, the ``long-wave'' radiation from the bunch moving along the planar periodic wire structure has been studied in papers~%
\cite{TVG14, TVG15}. 
Under this approximation, the structure is described by so-called averaged boundary conditions. 
Mentioned papers demonstrate that the bunch generates surface waves propagating along the wires, and the surface wave configuration can be used for determination the size of the bunch. 
However, despite the mentioned similarity, there is a principal difference between EBC and averaged boundary conditions, therefore the problem considered in the present paper is of independent value. 

The content of the paper is the following. 
First, we briefly recall the method of EBC. 
Next, we obtain the general exact solution of the problem and carry out the asymptotic analysis of this solution. 
The last section is devoted to the description of the main physical effect, which consists in generation of surface waves.
\section{Equivalent boundary conditions}
\begin{figure}[h]
	\includegraphics[width=\linewidth]{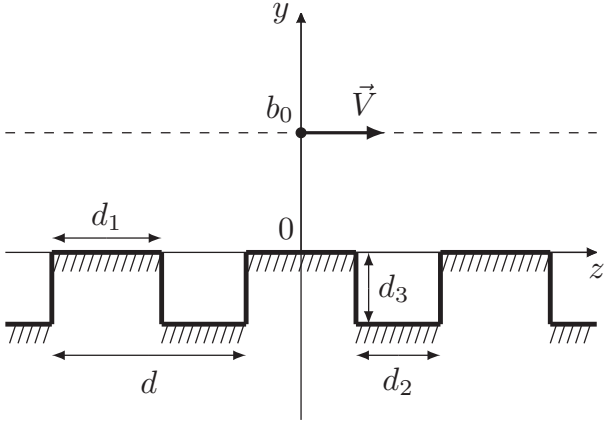}
	\caption{\label{fig1} The corrugated surface and moving charge.} 
\end{figure}

We consider a perfectly conductive flat surface having rectangular corrugation (Fig.~\ref{fig1}). 
It is assumed that the period $d$ and the depth of corrugation $d_3$ are much less than the wavelength under consideration $\lambda$:
\begin{equation}
\label{eq(2.1)}
\tag{2.1}
d\ll\lambda, \; d_3\ll\lambda.
\end{equation}
In this case, we can replace the corrugated surface with a planar surface, on which the so-called equivalent boundary conditions (EBC) are fulfilled~\cite{NefSivb77}. 
Note that the EBC are widely applied in the theory of corrugated structures (for instance, in~\cite{NefSivb77} one can find a lot of corresponding examples).
The EBC have the following general form for Fourier-transforms of electric and magnetic fields~\cite{NefSivb77}:
\begin{equation}
\label{eq(2.2)}
\tag{2.2}
E_{\omega z}=\eta^m H_{\omega x}, \; E_{\omega x}=\eta^e H_{\omega z},
\end{equation}
where $\eta^m$ and $\eta^e$ are ``impedances'', which are imaginary for perfectly conductive structures. In the case of the structure shown in Fig.~\ref{fig1}, we have~\cite{NefSivb77}:
\begin{equation}
\tag{2.3}
\label{eq(2.3)}
\eta^m=ik_0\left(d_0{-}\delta\frac{\alpha_z^2}{1{-}\alpha_x^2}\right)\!,\;\;\eta^e={-}ik_0\delta\left(1{-}\alpha_x^2\right)\!,
\end{equation}
where $d_0=d_2d_3/d$, $d_2$ is a width of groove, $\alpha_x=k_x/k_0$ and $\alpha_z=k_z/k_0$ are directional cosines of the incident wave with respect to $x$- and $z$-axis, correspondingly ($\mathbf{k}_0$ is the wave vector of the incident wave, $k_0=\omega/c$). 
The parameter of corrugation $\delta$ is determined by the formula~\cite{NefSivb77}:
\begin{align}
\label{eq(2.4)}
\delta&=d_3+\frac{d}{2\pi}\ln\left(\frac{\sigma-1}{\sigma}\right)+ \nonumber \\
\tag{2.4}
&+\frac{td}{2\pi}\int_{0}^{1/\sigma}\frac{du}{\sqrt{\left(1-u\right)\left(1-\sigma u\right)}\left(\sqrt{1-tu}+1\right)},
\end{align}
where parameters $t$ and $\sigma$ should be found from the following system of transcendent equations:
\begin{equation}
\label{eq(2.5)}
\tag{2.5}
\begin{split}
\int_{0}^{t}\frac{\sqrt{t-u}}{\sqrt{u\left(1-u\right)\left(\sigma-u\right)}}du&=\pi\frac{d_1}{d},\\
\int_{t}^{1}\frac{\sqrt{u-t}}{\sqrt{u\left(1-u\right)\left(\sigma-u\right)}}du&=2\pi\frac{d_3}{d}.
\end{split}
\end{equation}
Note that the parameter $\delta$ is of the same order as $d$ and $d_3$. 
In the specific case of a system of thin diaphragms (where $d_1/d\rightarrow0$) the parameter $\delta$ is determined by the following simple expression~\cite{NefSivb77}:
\begin{equation}
\label{eq(2.6)}
\tag{2.6}
\delta=d_3-\frac{d}{\pi}\ln\left[\cosh\left(\frac{\pi d_3}{d}\right)\right].
\end{equation}
\section{General solution of the problem}
We assume that the charged particle bunch moves along the corrugated surface with velocity $\mathbf{V}=c\beta\mathbf{e}_z$, which is perpendicular to corrugations. 
The distance between the charge and the surface is $b_0$ (Fig.~\ref{fig1}). 
The charge density is written in the form
\begin{equation}
\label{eq(3.1)}
\tag{3.1}
\rho=q\delta(x)\delta(y-b_0)\eta(z-Vt),
\end{equation}
where $\eta(z-Vt)$ is the charge distribution along the trajectory of the motion. 
We will use the Hertz potential $\mathbf{\Pi}$ and present it as a sum of an ``incident'' potential $\mathbf{\Pi}^{(i)}$ and a ``reflected'' potential $\mathbf{\Pi}^{(r)}$:
\begin{equation}
\label{eq(3.2)}
\tag{3.2}
\mathbf{\Pi}=\mathbf{\Pi}^{(i)}+\mathbf{\Pi}^{(r)}=\int_{-\infty}^{+\infty}\left(\mathbf{\Pi}_\omega^{(i)}+\mathbf{\Pi}_\omega^{(r)}\right)e^{-i\omega t}d\omega,
\end{equation}
where $\mathbf{\Pi}_\omega^{(i)}$ and $\mathbf{\Pi}_\omega^{(r)}$ are corresponding Fourier-transforms. 
We mean that the incident field is the field in unbounded vacuum, and reflected field is an additional field connected with the influence of the corrugated structure. 
Relations between the Fourier-transforms of electromagnetic field and Hertz potential are given by the formulae
\begin{equation}
\label{eq(3.3)}
\tag{3.3}
\mathbf{E}_\omega=\mathbf{\nabla}\operatorname{div}\mathbf{\Pi}_\omega+k_0^2\mathbf{\Pi}_\omega, \; \mathbf{H}_\omega=-ik_0\operatorname{rot}\mathbf{\Pi}_\omega.
\end{equation}
It is well known that from Maxwell's equations one can obtain the Helmholtz equation for the Hertz potential:
\begin{equation}
\label{eq(3.4)}
\tag{3.4}
\left(\Delta+k_0^2\right)\mathbf{\Pi}_\omega=-\frac{4\pi i}{ck_0}\mathbf{j}_\omega.
\end{equation}
The solution of Eq.~\eqref{eq(3.4)} for the incident field is the well-known Coulomb field of moving charge in unbounded vacuum. 
The Fourier-transform of the Hertz vector components of this field is
\begin{equation}
\label{eq(3.5)}
\tag{3.5}
\Pi_{\omega x}^{(i)}=\Pi_{\omega y}^{(i)}=0,
\end{equation}
\begin{align}
\label{eq(3.6)}
\Pi_{\omega z}^{(i)}=-\frac{q\tilde\eta}{ck_0}&\exp\left(i\frac{k_0z}{\beta}\right) \nonumber \\
\tag{3.6}
&\times\int_{-\infty}^{+\infty}dk_x\frac{e^{ik_xx+ik_{y0}|y-b_0|}}{k_{y0}},
\end{align}
where $k_{y0}=i\sqrt{k_x^2+k_0^2\frac{1-\beta^2}{\beta^2}}$, and $\operatorname{Im}k_{y0}>0$. 
Here $\tilde\eta$ is the Fourier-transform of the charge distribution along the trajectory of the motion $\eta(z-Vt)$:
\begin{equation}
\label{eq(3.7)}
\tag{3.7}
\tilde\eta=\frac{1}{2\pi}\int_{-\infty}^{+\infty}d\zeta\eta\left(\zeta\right)e^{-i\frac{k_0}{\beta}\zeta}, \;\;\zeta=z-vt.
\end{equation}
One can show that the reflected field is described by the two-component Hertz vector
\begin{equation}
\label{eq(3.8)}
\tag{3.8}
\mathbf{\Pi}_\omega^{(r)}=\Pi_{\omega x}^{(r)}\mathbf{e}_x+\Pi_{\omega z}^{(r)}\mathbf{e}_z,
\end{equation}
where
\begin{align}
\label{eq(3.9)}
\Pi_{\omega z}^{(r)}=-\frac{q\tilde\eta}{ck_0}&\exp\left(i\frac{k_0z}{\beta}\right) \nonumber \\
\tag{3.9}
&\times\int_{-\infty}^{+\infty}dk_xR_z\frac{e^{ik_xx+ik_{y0}(y+b_0)}}{k_{y0}},
\end{align}
\begin{align}
\label{eq(3.10)}
\Pi_{\omega x}^{(r)}=-\frac{q\tilde\eta}{ck_0}&\exp\left(i\frac{k_0z}{\beta}\right) \nonumber \\
\tag{3.10}
&\times\int_{-\infty}^{+\infty}dk_xR_x\frac{e^{ik_xx+ik_{y0}(y+b_0)}}{k_{y0}}.
\end{align}
Here, $R_z$ and $R_x$ can be called reflection coefficients, which should be found from the boundary conditions.

One can see that, in~\eqref{eq(3.6)} - \eqref{eq(3.10)}, the longitudinal component of the wave vector is $k_z=k_0/\beta$. 
Therefore, $\alpha_z=k_z/k_0=\beta^{-1}$, and Eq.~\eqref{eq(2.3)} is written in the form
\begin{equation}
\tag{3.11}
\label{eq(3.11)}
\eta^m{=}ik_0\!\left(\!d_0{-}\delta\frac{k_0^2}{\beta^2\!\left(k_0^2{-}k_x^2\right)}\right)\!,\;\;\eta^e{=}{-}i\frac{\delta}{k_0}\!\left(k_0^2{-}k_x^2\right)\!.
\end{equation}
Satisfying boundary conditions~\eqref{eq(2.2)} one can find the following expressions for the reflection coefficients:
\begin{widetext}
\begin{equation}
\label{eq(3.12)}
\tag{3.12}
R_z=-\frac{k_x^2k_0+\left(k_0-\beta^2k_0-k_{y0}\beta^2\eta^m\right)\left(k_0^2-k_x^2+k_{y0}k_0\eta^e\right)}{k_x^2k_0+\left(k_0-\beta^2k_0+k_{y0}\beta^2\eta^m\right)\left(k_0^2-k_x^2+k_{y0}k_0\eta^e\right)},
\end{equation}
\begin{equation}
\label{eq(3.13)}
\tag{3.13}
R_x=\frac{2k_0k_xk_{y0}\beta\eta^m}{k_x^2k_0+\left(k_0-\beta^2k_0+k_{y0}\beta^2\eta^m\right)\left(k_0^2-k_x^2+k_{y0}k_0\eta^e\right)},
\end{equation}
\end{widetext}
where $k_{y0}=i\sqrt{k_x^2+k_0^2\frac{1-\beta^2}{\beta^2}}$. 
Thus, we have found the general solution of the problem.

\section{Analysis of singularities in the Fourier-integrals}

Further, we will perform an asymptotic analysis of the obtained Fourier transforms of the field components. For simplicity, we will consider only the positive frequencies $\omega>0$. For negative frequencies, all formulas are easy to obtained by the rule $ F_{-\omega}=F^*_\omega $ where the asterisk means complex conjugation. This rule is true for Fourier transform of any real function $F(t)$, in particular, for the field components $ E_{x,y,z}(t)$ and $H_{x,y,z}(t)$. 

\begin{figure}[h]
	\includegraphics[width=\linewidth]{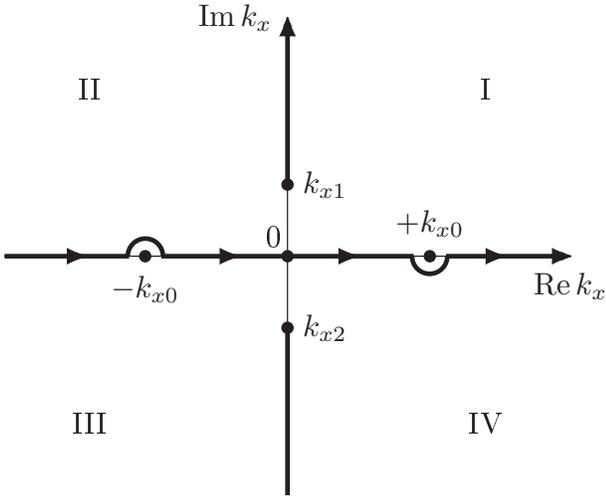}
	\caption{\label{fig2} The complex plane $k_x$: the integration path (real axis), the poles $\pm k_{x0}$, the branch points $k_{x1,2}$ and the cuts. The quadrants of the physical sheet of the complex plane $k_x$ are marked by Roman numerals.} 
\end{figure}

For our consideration, the poles of integrands in~\eqref{eq(3.9)} and~\eqref{eq(3.10)} will play an important role. They are found from the dispersion equation
\begin{equation}
\label{eq(4.1)}
\tag{4.1}
k_x^2k_0{+}\!\left(k_0{-}\beta^2k_0{+}k_{y0}\beta^2\eta^m\right)\!\left(k_0^2{-}k_x^2{+}k_{y0}k_0\eta^e\right)\!{=}0.
\end{equation}
Taking into account that $k_0d=2\pi d/\lambda$ and $k_0\delta$ are small parameters of the problem, one can find the following approximate solutions of Eq.~\eqref{eq(4.1)}:
\begin{equation}
\label{eq(4.2)}
\tag{4.2}
k_x=\pm k_{x0}=\pm\frac{k_0}{\beta}\sqrt{\frac{k_0^2}{\beta^2}\left(d_0-\frac{\delta}{\beta^2}\right)^2-\left(1-\beta^2\right)}.
\end{equation}
As we will see below, if these poles are real then their contributions are surface waves. 
Since the first term under radical is small then $k_{x0}$ can be real only for ultra-relativistic case ($1-\beta^2\ll 1$). 
Approximately, the condition of reality of~\eqref{eq(4.2)} has the form
\begin{equation}
\label{eq(4.3)}
\tag{4.3}
\beta>\sqrt{1-k_0^2\left(d_0-\delta\right)^2}\approx 1-\frac{k_0^2\left(d_0-\delta\right)^2}{2}.
\end{equation}
The integrands in~\eqref{eq(3.9)} and~\eqref{eq(3.10)} have as well other singularities: the function $k_{y0}=i\sqrt{k_x^2+k_0^2\frac{1-\beta^2}{\beta^2}}$ has two branch points $k_x=k_{x1,2}=\pm ik_0\frac{\sqrt{1-\beta^2}}{\beta}$. 
Fig.~\ref{fig2} shows the integration path, the poles $\pm k_{x0}$, the branch points $k_{x1,2}$ and the cuts in the complex plane $k_x$. 
Taking into account some negligible small losses (i.e., negligible small imaginary part of $k_0$) one can show that the poles $\pm k_{x0}$ are located with respect to the integration path as shown in Fig.~\ref{fig2}.

\begin{figure}[h]
	\includegraphics[width=\linewidth]{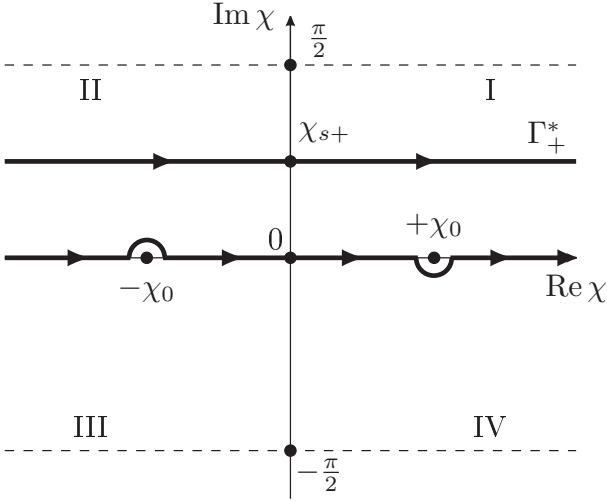}
	\caption{\label{fig3} The complex plane $\chi$: initial integration path, the poles $\pm\chi_0$, the saddle point $\chi_{s+}$ and the steepest descent path $\Gamma_+^*$ (for $x>0$). Roman numerals designate the areas, which correspond to the quadrants of the physical sheet of the complex plane $k_x$.} 
\end{figure}

To investigate an asymptotic behavior of Fourier-integrals~\eqref{eq(3.9)} and~\eqref{eq(3.10)} we can use the saddle point method~%
\cite{FMb, BHb}. 
First, it is convenient to change the integration variable according to the rule
\begin{equation}
\label{eq(4.4)}
\tag{4.4}
k_x=\frac{k_0}{\beta}\sqrt{1-\beta^2}\operatorname{sh}\chi.
\end{equation}
Then the phase $\Phi\left(k_x\right)=ik_{y0}\left(y+b_0\right)+ik_xx$ is written in the following form:
\begin{equation}
\label{eq(4.5)}
\tag{4.5}
\Phi\left(\chi\right)=\frac{k_0}{\beta}\sqrt{1-\beta^2}\left[x\operatorname{sh}\chi+i\left(y+b_0\right)\operatorname{ch}\chi\right].
\end{equation}
Saddle points are found from the equation
\begin{equation}
\label{eq(4.6)}
\tag{4.6}
\frac{d\Phi\left(\chi\right)}{d\chi}=\frac{k_0}{\beta}\sqrt{1-\beta^2}\left[x\operatorname{ch}\chi+i\left(y{+}b_0\right)\operatorname{sh}\chi\right]=0.
\end{equation}
It gives two saddle points $\operatorname{ch}\chi_{s\pm}=\pm\frac{y+b_0}{\sqrt{x^2+\left(y+b_0\right)^2}}$, which are purely imaginary: $\chi_{s\pm}=i\chi_{s\pm}^{\prime\prime}$. 
According to~\eqref{eq(4.4)}, mapping of the physical sheet of the complex plane $k_x$ is the strip $\left(-\infty<\operatorname{Re}\chi<+\infty,-\pi/2<\operatorname{Im}\chi<\pi/2\right)$ on the complex plane $\chi$ (Fig.~\ref{fig3}). 
There is only one saddle point $\chi_{s+}$ in this area. 
The steepest descent path $\Gamma_+^*$ contains the point $\chi_{s+}$ and can be found from the system
\begin{equation}
\tag{4.7}
\label{eq(4.7)}
\begin{split}
	&\operatorname{Re}\Phi\left(\chi\right)=\operatorname{Re}\Phi\left(\chi_{s+}\right),\\
	&\operatorname{Im}\Phi\left(\chi\right)>\operatorname{Im}\Phi\left(\chi_{s+}\right).
\end{split}
\end{equation}
According to Eqs.~\eqref{eq(4.7)} the steepest descent path $\Gamma_+^*$ is a straight line parallel to the real axis (Fig.~\ref{fig3}).

The initial integration path can be transformed to the path $\Gamma_+^*$ with separation of the contribution of the pole $+k_{x0}$ for $x>0$ (or $-k_{x0}$ for $x<0$) that is
\begin{equation}
\tag{4.8}
\label{eq(4.8)}
\int_{-\infty}^{+\infty}=\int\limits_{\Gamma_+^*}+2\pi i\underset{k_x=k_{x0}}{\operatorname{Res}}.
\end{equation}
The integrals over the steepest descent path can be asymptotically estimated by the well-known formula (see the books~%
\cite{FMb, BHb}%
):
\begin{align}
\label{eq(4.9)}
\left\{\Pi_{\omega z}^{(r)} \atop \Pi_{\omega x}^{(r)}\right\}& \approx i\frac{q\tilde\eta}{ck_0}\left(\frac{2\pi\beta}{k_0\sqrt{1-\beta^2}\sqrt{x^2+\left(y+b_0\right)^2}}\right)^{\frac{1}{2}} \nonumber \\
\times&\left\{R_z\left(\chi_{s+}\right) \atop R_x\left(\chi_{s+}\right)\right\}\operatorname{exp}\left(\frac{ik_0z}{\beta}\right) \nonumber \\
\tag{4.9}
&\times\operatorname{exp}\left(-\frac{k_0}{\beta}\sqrt{1-\beta^2}\sqrt{x^2+\left(y+b_0\right)^2}\right)\!.
\end{align}

Formula~\eqref{eq(4.9)} demonstrates that the contribution of the saddle point decreases exponentially with increase in $|x|$ and $|y|$. 
We can neglect this contribution if the following inequality is satisfied:
\begin{equation}
\tag{4.10}
\label{eq(4.10)}
\frac{k_0}{\beta}\sqrt{1-\beta^2}\sqrt{x^2+\left(y+b_0\right)^2}\gg 1.
\end{equation}

Further, we perform an asymptotic analysis of the field components assuming that $|x|$ and $|y|$ are sufficiently great to satisfy inequality~\eqref{eq(4.10)}. 
In this case, we can neglect the contribution of the saddle point, and Fourier integrals~\eqref{eq(3.9)} and~\eqref{eq(3.10)} are determined  by the contribution of the pole $+k_{x0}$ only (for $x>0$; or $-k_{x0}$ for $x<0$). 
Underline that exponential decrease of the saddle point contribution means the absence of volume radiation.

The absence of volume radiation can be explained by the following way. 
The field of the charge on the frequency under consideration must be propagated  along the charge trajectory with the phase velocity being equal to the charge velocity (like Cherenkov radiation in the material media). 
Therefore, at the frequency $\omega$, we have $k_z=\omega/V>\omega/c$ , and the dispersion equation results in $k_x^2+k_y^2=\omega^2/c^2-\omega^2/V^2$. 
This equation has no solution if $k_x$  and $k_y$ are real. 
This predetermines the absence of volume radiation. 

It should be underlined that this conclusion refers to the range of relatively low frequencies when the wavelength is much more than the structure period, and we can describe the real structure with help of the EBC. Well-known Smith-Purcell radiation is not generated on these frequencies. 
Smith-Purcell radiation is an effect of superposition of diffraction radiation from different periods of the structure. 
The phenomena considered here are not connected with diffraction radiation; they are determined by specific anisotropy of the structure.

\section{Surface waves}
The contribution of the pole $+k_{x0}$ for $x>0$ (and $-k_{x0}$ for $x<0$) is a surface wave under condition that the pole is real, i.e. when inequality~\eqref{eq(4.3)} is satisfied. 
Accordingly to~\eqref{eq(4.3)}, in this case the bunch is ultra-relativistic. 
Therefore, further we will assume that the parameter $\beta$ is close to unity ($\beta\rightarrow 1$). Taking into account this approximation, we can write expressions for poles~\eqref{eq(4.2)} in the following way:
\begin{equation}
\tag{5.1}
\label{eq(5.1)}
\pm k_{x0}=\pm k_0^2|d_0-\delta|.
\end{equation}
Calculating the residues we obtain for the Hertz vector of the surface waves
\begin{align}
\label{eq(5.2)}
\left\{\Pi_{\omega z}^{(s)} \atop \Pi_{\omega x}^{(s)}\right\}&=2\pi i\operatorname{sgn}\left(x\right)\underset{k_x=\pm k_{x0}}{\operatorname{Res}}\left\{\Pi_{\omega z}^{(r)} \atop \Pi_{\omega x}^{(r)}\right\}= \nonumber \\
&=4\pi\frac{q\tilde\eta}{c}\left\{k_0^{-1} \atop \operatorname{sgn}\left(x\right)|d_0-\delta|\right\}\operatorname{exp}\left(ik_0z\right) \nonumber \\
\tag{5.2}
&\times\operatorname{exp}\!\left[ik_0^2|d_0{-}\delta||x|{-}k_0^2|d_0{-}\delta|\left(y{+}b_0\right)\right]\!.
\end{align}
The corresponding field components are
\begin{align}
\label{eq(5.3)}
\left\{\begin{aligned}
E_{\omega x}^{(s)} \\
E_{\omega y}^{(s)} \\
E_{\omega z}^{(s)}\end{aligned}\right\}
&=4\pi i\frac{q\tilde\eta k_0}{c}
\left\{\begin{aligned}
&\ \ 0 \\
-k_0&|d_0-\delta| \\
ik_0^2&\left(d_0-\delta\right)^2\end{aligned}\right\}\operatorname{exp}\left(ik_0z\right) \nonumber \\
\tag{5.3}
&\times\!\operatorname{exp}\left[ik_0^2|d_0{-}\delta||x|{-}k_0^2|d_0{-}\delta|\left(y{+}b_0\right)\right]\!,
\end{align}
\begin{align}
\label{eq(5.4)}
\left\{\begin{aligned}
H_{\omega x}^{(s)} \\
H_{\omega y}^{(s)} \\
H_{\omega z}^{(s)}\end{aligned}\right\}
&{=}4\pi i\frac{q\tilde\eta k_0}{c}\!
\left\{\begin{aligned}
&k_0|d_0{-}\delta| \\
&\ \ \ \ \ 0 \\
{-}\operatorname{sgn}&\left(x\right)\!k_0^2\left(d_0{-}\delta\right)^2\end{aligned}\right\}\!\operatorname{exp}\left(ik_0z\right) \nonumber \\
\tag{5.4}
&\times\!\operatorname{exp}\left[ik_0^2|d_0{-}\delta||x|{-}k_0^2|d_0{-}\delta|\left(y{+}b_0\right)\right]\!.
\end{align}
These results are true for $k_0=\omega/c>0$. Recall that corresponding formulae for negative frequencies are obtained by the rule $F_{-\omega}=F^*(\omega)$. 

Note that, in ~\eqref{eq(5.3)}, \eqref{eq(5.4)}, we take into account only the main terms with respect to the small parameter $k_0|d_0-\delta|$. 
Note that the components $E_{\omega x}^{(s)}$ and $H_{\omega y}^{(s)}$ vanish in the approximation under consideration (we can neglect these components because they are determined by terms of third order of smallness). 
As we see, all components decrease exponentially with an increase in $y$, i.e. formulae~\eqref{eq(5.3)} --~\eqref{eq(5.4)} describe the surface waves propagating in a plane of the structure.

It is also of interest to analyze energy losses of the bunch, which can be obtained by calculation of the energy flow through two parallel half-planes $x=\pm x_0$, $y>0$. This  way results in the following expression for the energy losses per unit of the path length:
\begin{equation}
\tag{5.5}
\label{eq(5.5)}
\frac{dW}{dz_0}=\frac{1}{V}\frac{dW}{dt}=\frac{2}{c\beta}\int_{-\infty}^{+\infty}dz\int_{0}^{+\infty}dyS_x\bigg|_{x=x_0>0},
\end{equation}
where $S_x$ is the $x$-component of the the energy flow density  
\begin{equation}
\tag{5.6}
\label{eq(5.6)}
S_x=\frac{c}{4\pi}\left(E_y^{(s)}H_z^{(s)}-E_z^{(s)}H_y^{(s)}\right).
\end{equation}
The factor ``2'' in~\eqref{eq(5.5)} is explained by generation of two symmetrical surface waves in positive and negative directions of $x$. 

\begin{figure*}[htb]
	\begin{center}
		\begin{minipage}[t]{0.48\linewidth}
			\includegraphics[width=1.1\linewidth]{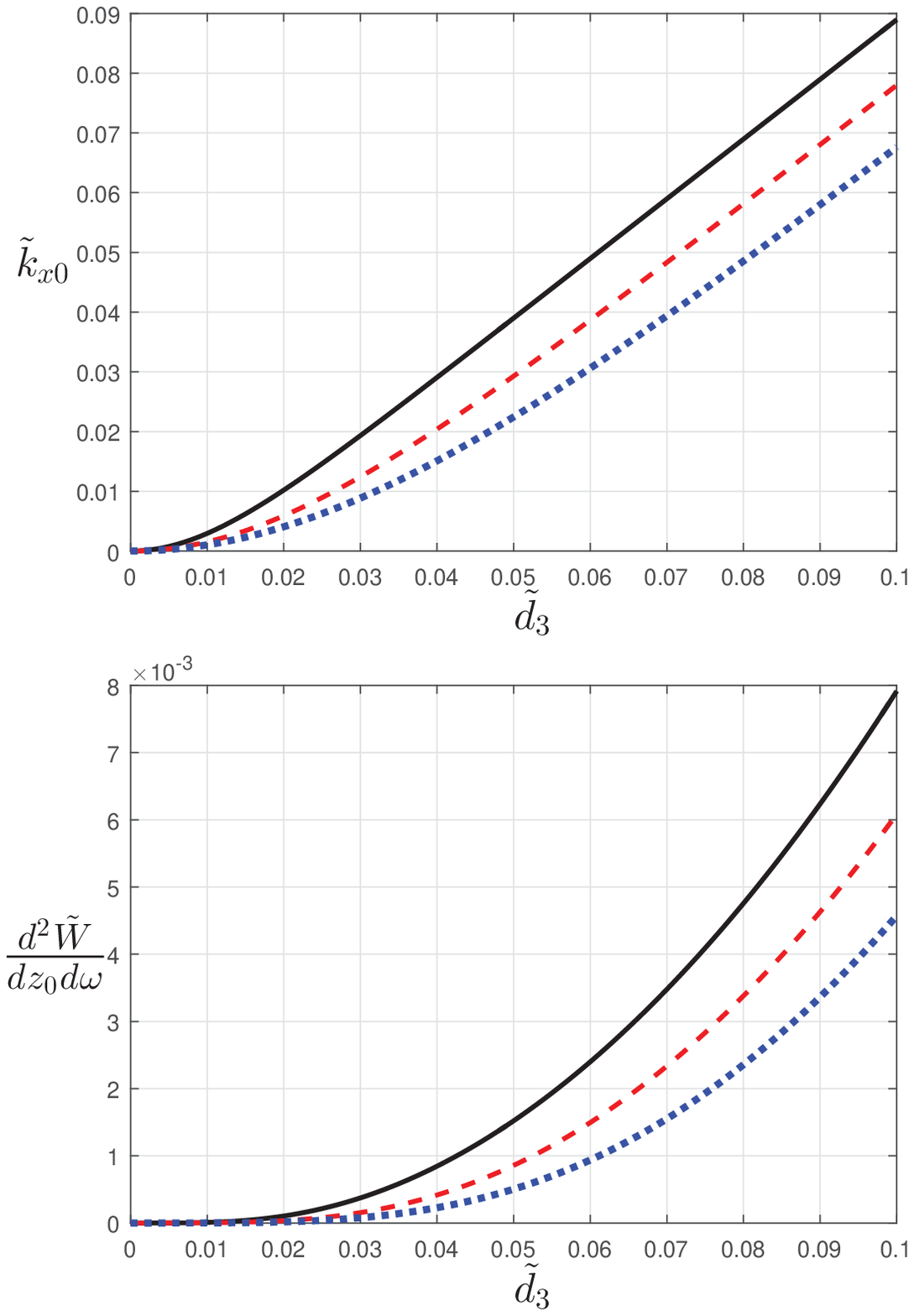}
			\caption{\label{fig4} The wave vector projection $\tilde k_{x0}$ and the spectral density of energy losses $d^2\tilde W/d\omega dz_0$ depending on the depth of corrugations $\tilde d_3$ for the following parameters: $\beta \rightarrow 1$, $b_0=0$, $d_1 \rightarrow 0$, $d_2 \rightarrow d$; $d=0.05$ (solid black curves); $d=0.1$ (dashed red curves); $d=0.15$ (dotted blue curves). All parameters of corrugated structure are in units $k_0^{-1}$.}
		\end{minipage}
		\hspace{0.005\linewidth}
		\begin{minipage}[t]{0.48\linewidth}
			\includegraphics[width=1.1\linewidth]{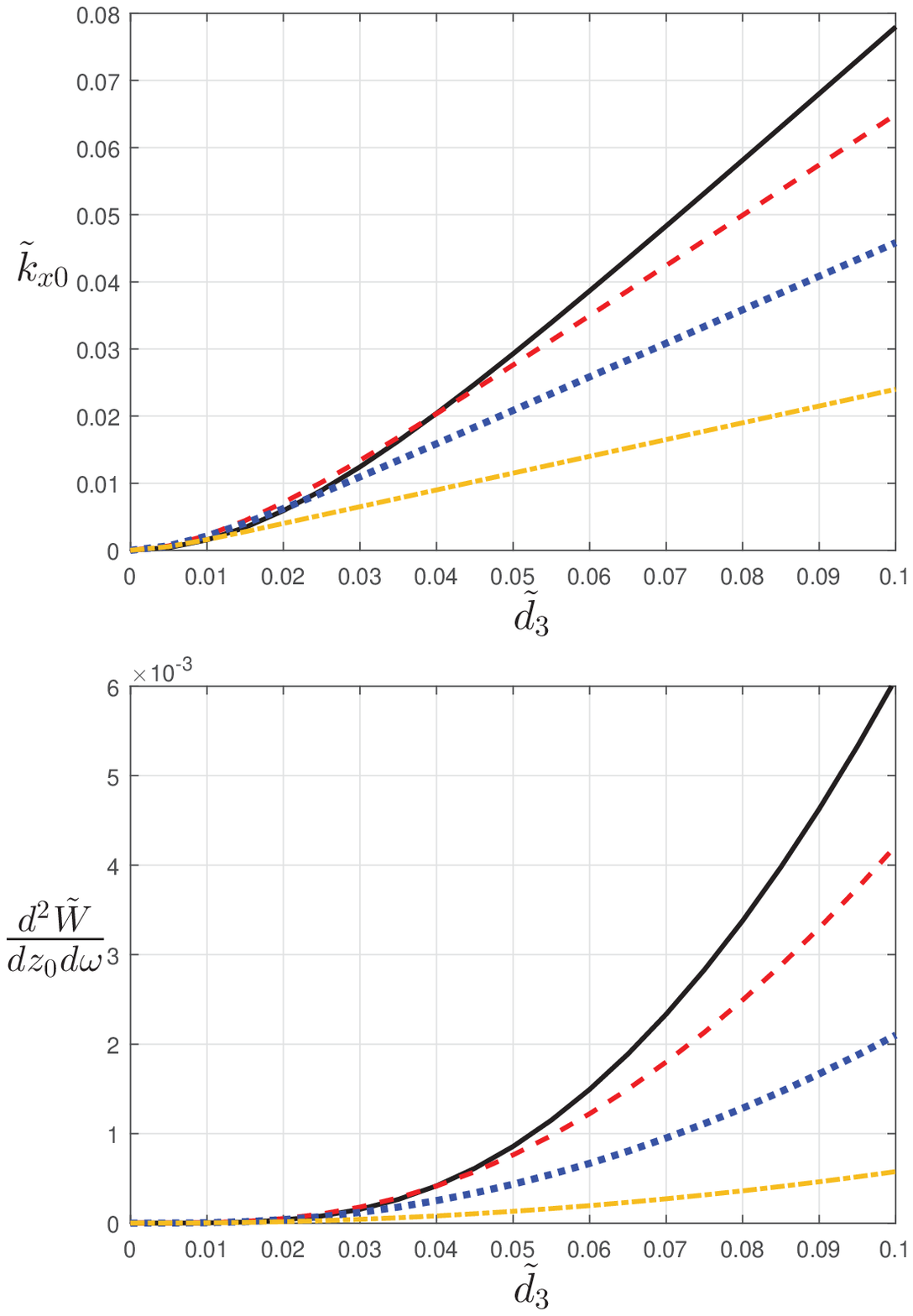}
			\caption{\label{fig5} The wave vector projection $\tilde k_{x0}$ and the spectral density of energy losses $d^2\tilde W/d\omega dz_0$ depending on the depth of corrugations $\tilde d_3$ for the following parameters: $\beta \rightarrow 1$, $b_0=0$, $d=0.1$; $d_1 \rightarrow 0$, $d_2 \rightarrow d$ (system of diaphragms, solid black curves); $d_1=0.025$, $d_2=0.075$, (dashed red curves); $d_1=d_2=0.05$ (dotted blue curves); $d_1=0.075$, $d_2=0.025$ (dashed-dotted yellow curves). All parameters of corrugated structure are in units $k_0^{-1}$.}
		\end{minipage}
	\end{center}
\end{figure*}

Writing components of electromagnetic field as the Fourier-integrals $F\left(\vec{r},t\right)=\int_{-\infty}^{+\infty}d\omega F_\omega e^{-i\omega t}$ and substituting these integrals into~\eqref{eq(5.5)}, \eqref{eq(5.6)} we obtain after some transformations the following integral over frequency:
\begin{equation}
\tag{5.7}
\label{eq(5.7)}
\frac{dW}{dz_0}=\int_{0}^{+\infty}\frac{d^2W}{d\omega dz_0}d\omega,
\end{equation}
where $d^2W/d\omega dz_0$ is a spectral density of the energy losses
\begin{equation}
\tag{5.8}
\label{eq(5.8)}
\frac{d^2W}{d\omega dz_0}=2c\!\int_{0}^{+\infty}\!dy\operatorname{Re}\!\left(E_{\omega y}^{(s)}H_{\omega z}^{(s)^*}-E_{\omega z}^{(s)}H_{\omega y}^{(s)^*}\right).
\end{equation}

Substituting formulae~\eqref{eq(5.3)} and~\eqref{eq(5.4)} into~\eqref{eq(5.8)} and performing the integration one can obtain the following expression:
\begin{equation}
\tag{5.9}
\label{eq(5.9)}
\frac{d^2W}{d\omega dz_0}=16\pi^2\frac{q^2|\tilde\eta|^2k_0}{c}\frac{d^2\tilde W}{d\omega dz_0},
\end{equation}
where
\begin{equation}
\tag{5.10}
\label{eq(5.10)}
\frac{d^2\tilde W}{d\omega dz_0}=k_0^2\left(d_0-\delta\right)^2e^{-2k_0^2|d_0-\delta|b_0}
\end{equation}
is dimensionless spectral density of the energy losses. 
Further, we use as well the dimensionless projection of the wave vector:
\begin{equation}
\tag{5.11}
\label{eq(5.11)}
\tilde k_{x0}=\frac{k_{x0}}{k_0}=k_0|d_0-\delta|.
\end{equation}

We demonstrate below some results of computations performed using the formulas given above. These computations are performed for the conditions providing that the EBC technique is obviously applicable. The small parameter of the problem 
$d/\lambda \approx d k_0 / (2 \pi)$
is from $0.05 / (2 \pi)$  to $0.15 / (2 \pi)$  
for the situations considered below. Meanwhile, earlier we have tested the EBC technique for the case of a circular corrugated waveguide with small parameter 
$d/\lambda \sim 0.1$, 
and it has been shown that even such moderately small parameter provides the accuracy (compared to CST simulated results) within several percents \cite{TVAA2018}.

Fig.~\ref{fig4} shows dependences of the wave vector projection and the spectral density of the energy losses on the depth of corrugations $d_3$ for different periods of the corrugated surface. 
It is assumed that $d_1\rightarrow 0$, $d_2\rightarrow d$, $d_0=d_2d_3/d\rightarrow d_3$ (system of diaphragms). 
According to the plots, the $x$-component of the wave vector $\tilde k_{x0}$ increases with increasing the depth of corrugations $d_3$, and this dependence is almost linear in the most part of the plot. 
This is explained by the fact that the parameter $\delta$ only weakly depends on the depth for not small values of $d_3/d$ (then, accordingly to \eqref{eq(5.1)}, the function $k_{x0}\left(d_3\right)$ is almost linear). 
Energy losses increase approximately with the square of the depth of corrugations. 
It should be also emphasized that both the projection of the wave vector and the spectral density of the energy losses decrease with increasing the period $d$.

Fig.~\ref{fig5} shows dependences of the wave vector projection and the spectral density of the energy losses on the depth of corrugations $d_3$ for different values of parameters $d_1$ and $d_2$ (for the same period $d$). 
One can see that the  $x$-component of the wave vector and the spectral density of the energy losses usually increase with decrease in the width of ``hills'' $d_1$, and the system of diaphragms is the most effective for the surface waves generation. 
However, if the depth of corrugations $d_3$ is small enough then this regularity can be broken.

Concluding this section, we emphasize that the radiation under consideration is fundamentally different from the Smith-Purcell one. Unlike Smith-Purcell radiation, the radiation considered here is surface, has a continuous spectrum and is generated at lower frequencies. However, there are some common features. In particular, the  energy of radiation of both types is proportional to the square of the charge and the structure length, and also significantly depends on the geometrical parameters of the cell of the structure.

\section{Conclusion}

We have investigated radiation of the charged particle bunch moving along the corrugated metallic surface. 
The principal point here is that we consider radiation at the wavelengths, which are much larger compared to the structure periods (in contrast to traditional consideration for wavelengths which are comparable with the period). 
The general solution of the problem has been obtained by use of the method of the equivalent boundary conditions. 
We have performed asymptotic analysis of the electromagnetic field and have shown that the volume radiation is absent in the situation under consideration but the surface waves can be excited. 
As a rule, these waves are generated if the charge velocity is close to the light velocity. 
Analysis of the surface waves has demonstrated the main dependences of the energy losses on the problem parameters. 
In particular, it has been shown that the energy losses increase with increase in the depth of corrugations and, usually, as well with decrease in the width of ``hills''.


\begin{acknowledgments}

This work was supported by the Russian Science Foundation (Grant No.~18-72-10137).

\end{acknowledgments}


%

\end{document}